\documentclass[12pt,letterpaper]{article}
\pdfoutput=1

\usepackage{graphicx}
\usepackage[body={17.5cm, 22cm},right=2cm]{geometry}
\usepackage{amsmath,amsthm,amsfonts,amssymb,amscd}
\usepackage{cite}

\usepackage{colortbl}
\definecolor{darkgreen}{RGB}{50,150,0}

\definecolor{darkblue}{rgb}{0.15,0.35,0.55}
\definecolor{reddish}{rgb}{0.65, 0.2, 0.2}

\usepackage[linktocpage=true]{hyperref}
\hypersetup{
colorlinks=true,
citecolor=darkblue,
linkcolor=reddish,
urlcolor=darkblue,
pdfauthor={},
pdftitle={},
pdfsubject={}
}

\newcommand\be{\begin{equation}}
\newcommand\ee{\end{equation}}
\newcommand\bea{\begin{eqnarray}}
\newcommand\eea{\end{eqnarray}}
\newcommand{\sfrac}[2]{{\textstyle\frac{#1}{#2}}}
\newcommand\di{\partial}

\newcommand\e{{\rm e}}
\newcommand\dd{{\rm d}}
 
\newcommand\msbar{\overline{\rm MS}}

\begin{document}

\thispagestyle{empty}
\begin{center}
{
\Large \bf Integrating out beyond tree level\\[.3cm]
and relativistic superfluids} \\[.8cm]
\large{Austin Joyce,$^{\rm a}$ Alberto Nicolis,$^{\rm b}$ Alessandro Podo,$^{\rm b}$  and Luca Santoni$^{\rm c}$}
\\[0.4cm]

\vspace{.2cm}
\small{\textit{$^{\rm a}$Kavli Institute for Cosmological Physics, Department of Astronomy and Astrophysics,  \\
The University of Chicago, Chicago, IL 60637, USA,}}

\vspace{.2cm}
\small{\textit{$^{\rm b}$Center for Theoretical Physics and Department of Physics, \\
  Columbia University, New York, NY 10027, USA}}
  
\vspace{.2cm}
\small{\textit{$^{\rm c}$ICTP, International Centre for Theoretical Physics,  \\
  Strada Costiera 11, 34151, Trieste, Italy}}

\end{center}

\vspace{.2cm}

\begin{abstract}
\noindent
We revisit certain subtleties of renormalization that arise when one derives a low-energy effective action by integrating out the heavy fields of a more complete theory. Usually these subtleties are circumvented by matching some physical observables, such as scattering amplitudes, but a more involved procedure is required if one is interested in deriving the effective theory to all orders in the light fields (but still to fixed order in the derivative expansion). As a concrete example, we study the $U(1)$ Goldstone low-energy effective theory that describes the spontaneously broken phase of a $\phi^4$ theory for a complex scalar. Working to lowest order in the derivative expansion, but to all orders in the Goldstones, we integrate out the radial mode at one loop and express the low-energy effective action in terms of the renormalized couplings of the UV completion. This yields the one-loop equation of state for the superfluid phase of (complex) $\phi^4$. We perform the same analysis for a renormalizable scalar $SO(N)$ theory at finite chemical potential, integrating out the gapped Goldstones as well, and confirm that the effective theory for the gapless Goldstone exhibits no obvious sign of the original $SO(N)$ symmetry. 
\end{abstract}

\newpage
\section{Introduction}
Within the path-integral formulation of quantum field theory, a low-energy effective field theory can be defined through the procedure of {\it  integrating out}. Concretely, if one is given a fundamental theory involving heavy fields $h(x)$ and light fields $\ell(x)$ which is described by an action $S[h,\ell \,]$, the low-energy effective action $S_{\rm eff}[\ell \,]$---for the light fields only---is formally defined by integrating out the heavy fields:
\be \label{definition}
\e^{i S_{\rm eff}[\ell \, ]} \equiv \int \!  Dh \, \e^{i S[h, \ell \,]} \; .
\ee
To the extent to which path-integral manipulations make sense, this definition is concrete and predictive. In particular, one can compute correlation functions of the light fields by performing a path integral over them weighted by the exponential of the effective action:
\be \label{correlators}
\langle \ell(x_1) \dots \ell(x_n) \rangle = \frac{\displaystyle{\int} \!  D\ell Dh  \, \e^{i S[h, \ell \,]}\, \ell(x_1) \cdots \ell(x_n)}{\displaystyle{\int} \!  D\ell Dh  \, \e^{i S[h, \ell \,]}} = \frac{\displaystyle{\int} \!  D\ell   \, \e^{i S_{\rm eff}[\ell \,]} \,\ell(x_1) \cdots \ell(x_n)}{\displaystyle{\int} \!  D\ell  \, \e^{i S_{\rm eff}[\ell \,]}}\,.
\ee
In fact, at this level there is no low-energy expansion whatsoever. The only restriction is that, not surprisingly, one uses the effective action for $\ell$ to compute correlation functions of $\ell$ only, but not of $h$. Put another way: as long as one is only interested in correlation functions of $\ell$, one can perform the path-integral over $h$ once and for all, which is what eq.~\eqref{definition} does.

\vskip4pt
Importantly, all this continues to be true beyond tree level, where one starts encountering ultraviolet (UV) divergences. Provided that the fundamental theory is regulated, and the necessary counterterms to renormalize it are included in $S[h, \ell \, ]$, 
eq.~\eqref{definition} yields the  action $S_{\rm eff}[\ell \,]$ for the regulated effective theory, including the necessary counterterms to renormalize it. This is a consequence of eq.~\eqref{correlators}; as far as correlation functions of the light fields go, the effective theory is just an exact rewriting of the full theory.

\vskip4pt
The power of effective field theory lies in the derivative expansion, truncated at some finite order. From the bottom-up perspective, this is a reflection of several dearly held principles: 
 locality, perturbativity, predictivity, {\it et cetera}. From the top-down perspective of eq.~\eqref{definition}, if one actually wants to derive the effective action from the fundamental theory, a pragmatic motivation for  the derivative expansion is that, in general, one cannot perform the path integral in \eqref{definition} for arbitrary field configurations $\ell(x)$, even working perturbatively in the couplings of the fundamental action $S[h, \ell \, ]$. 

\vskip4pt
However, at this point a problem arises: the derivative expansion completely changes the high-energy behavior of the theory, and thus the structure of UV divergences. As a result, the counterterms necessary to renormalize the effective theory correctly---so that it reproduces the same results as the full theory, when appropriate---are {\em not} those that come out of \eqref{definition}. In the  simplest terms, the problem has to do with expanding the propagators of the heavy fields in momentum, which is one way to think about the derivative expansion:
\be
\frac{i}{q^2 -M^2 + i\epsilon} \quad \xrightarrow{q^2\ll M^2} \quad -\frac{i}{M^2} \left[ 1 + \frac{q^2}{M^2} + \frac{q^4}{M^4} + \dots \right]\,.
\ee
The left-hand side decays at large $q$, while on the right-hand side the first term is a constant, and all the others {\em increase} at large $q$, faster and faster as we push the derivative expansion further. Clearly, loop integrals will have completely different UV properties depending on whether or not the propagator is expanded~\cite{Georgi:1993mps}. (For an explicit example of this at work, see~\cite{Penco:2020kvy}.)

\vskip4pt
So, if we work with the effective theory to some finite order in the derivative expansion, it seems that eq.~\eqref{definition} has {\em no} predictive power beyond tree-level. To renormalize the effective theory in a manner consistent with the full theory---so that the two agree for low-energy observables---we have to introduce new counterterms, because those that come out of \eqref{definition} are not the correct ones. But then, we face the following difficulty:  what comes out of the right-hand side of~\eqref{definition}, once expanded in derivatives, is {\em a} local functional of the light fields compatible with the symmetries of the theory, whereas the new counterterms we have to introduce are  {\em the most general} local terms compatible with the symmetries. Functionally, this makes eq.~\eqref{definition} useless. 

\vskip4pt
This is one of the reasons why, beyond tree level, eq.~\eqref{definition} is usually bypassed altogether. Instead of using \eqref{definition}, one parametrizes the most general effective theory compatible with the symmetries to the desired order in the derivative expansion, computes as many low-energy observables as needed to fix all the parameters in the effective theory, and matches them to the same observables computed in the full theory, thereby fixing the effective theory's parameters in terms of those of the full theory. This procedure is called ``matching."
This works well in practice for a finite number of parameters, which is usually all we need in particle physics when we are after, say, a low-energy amplitude with a fixed number of external legs at some finite order in the loop expansion. In that case, we are only interested in the effective theory to finite order in both the derivative expansion {\em and} in the number of fields.

\vskip4pt
However, there are other cases---most notably a relativistic field theory at finite density---where we might be interested in an effective theory to fixed order in the derivative expansion, but to {\em all} orders in the number of fields. What are we to do then? Should we match infinitely many amplitudes? Moreover, it is a pity that such a simple and beautiful definition as eq.~\eqref{definition} loses its operational meaning beyond tree level, becoming essentially a cartoon of the whole story. Can one salvage it, perhaps by modifying it?

\vskip4pt
Georgi~\cite{Georgi:1993mps} has proposed a formal matching procedure, whose definition does not require expanding in the number of fields. The basic idea is to use the quantum effective action for the light fields, $\Gamma[\ell \,]$. All correlation functions of the light fields can be computed directly using $\Gamma[\ell \,]$ at tree level, and so there are no more subtleties regarding UV divergences, even upon expanding in derivatives. Thus, the all-orders matching requirement is simply that the effective action $S_{\rm eff}[\ell \,]$ be such that the {\em quantum} effective action for $\ell$ computed in the effective theory is the same as that computed in the full theory:
\be
\Gamma_{\rm EFT}[\ell \,] = \Gamma_{\rm full}[\ell \, ] \equiv \Gamma [\ell \,] \; .
\ee
At the path integral level, this corresponds to modifying \eqref{definition} to read
\be
\bigg(\int  D \delta \ell \, \e^{i S_{\rm eff}[\ell+\delta \ell] }\bigg)_{\rm 1P I}  =
\bigg(\int Dh \, D \delta \ell \, \e^{i S[h, \, \ell+\delta \ell] }\bigg)_{\rm 1P I \; for \; \delta \ell} \; .
\ee
On the left-hand side, the path integral is restricted to one-particle-irreducible (1PI) diagrams, and simply computes the quantum effective action in the effective theory. On the right-hand side, the path integral is restricted to diagrams that are 1PI for the fluctuations of $\ell$, while $h$ is integrated out completely. This computes the quantum effective action for $\ell$ only, but in the full theory.\footnote{A somewhat different alternative to matching has been proposed in \cite{Fuentes-Martin:2016uol}.}

\vskip4pt
This definition of $S_{\rm eff}[\ell \,]$ is somewhat implicit. One should still parametrize $S_{\rm eff}[\ell \,]$ as the most general local effective action compatible with the symmetries. Then, if one is able to compute the quantum effective action both in the effective theory and in the full theory, one can fix the free parameters in $S_{\rm eff}[\ell \,]$. To the best of our knowledge, this procedure has not been attempted in practice. 
Our aim in this paper is to carry it out at one loop in two simple cases, which are relevant for relativistic field theories at finite density---in particular, for relativistic superfluids.

\vskip4pt
Of course, in general we are not able to compute quantum effective actions for generic field configurations, even stopping at one loop order. However, the procedure is systematic within the derivative expansion. In particular, to zeroth order in derivatives, it amounts to computing the Coleman--Weinberg potential \cite{Coleman:1973jx}, which is quite easy to do with functional methods \cite{Weinberg:1996kr}. To higher orders, the computation in general becomes substantially more involved \cite{Cheyette:1985ue, Chan:1986jq}, but such complications will not matter for us. In the two examples that we will study below, by applying essentially the same technique as used for the Coleman--Weinberg potential, we will be able to carry out the computation to all orders in a Goldstone field, and to the order of one derivative {\em per field} in the derivative expansion.

\vspace{.5cm}
\noindent
{\bf Notation and conventions:} We use natural units and the mostly minus signature for the metric throughout.
We will overuse the symbol $\mu$: as a Lorentz index, as a chemical potential, and as a renormalization scale (though typically we will denote the renormalization scale as $\bar\mu$ in $\msbar$). Unfortunately, in all of these cases $\mu$ is the conventional symbol of choice, hopefully the meaning will be clear from the context.

\section{Weakly coupled UV completions for relativistic superfluids}\label{UV completions}

From a quantum field theory (QFT) perspective, a superfluid can be defined as  a system with a spontaneously broken internal $U(1)$ symmetry, in a state of finite density (or finite chemical potential) for the associated charge \cite{Son:2002zn, Nicolis:2011pv}. Its low-energy effective theory involves a derivatively coupled Goldstone field $\pi(x)$, the superfluid phonon, on which the original $U(1)$ symmetry acts as a shift symmetry,
\be
\pi \mapsto \pi +{\rm const}.
\ee
The symmetry breaking pattern is conveniently parametrized in terms of a spacetime scalar $\phi(x)$ with a time-dependent expectation value. The phonon $\pi(x)$ describes fluctuations about this expectation value \cite{Nicolis:2013lma}:
\be
\phi(x) = \mu t + \pi(x) \; ,
\ee
where $\mu$ is the chemical potential. Notice that $\pi(x)$ nonlinearly realizes not only the internal $U(1)$ symmetry, but also time translations and Lorentz boosts.\footnote{That a single Goldstone field is responsible for nonlinearly realizing more than one spacetime symmetry is a manifestation of the so-called inverse Higgs effect~\cite{Ivanov:1975zq}.} (Though a diagonal combination of time translations and internal shifts is linearly realized.)

\vskip4pt
To lowest order in derivatives, the low-energy effective action is simply
\be \label{P(X)}
S_{\rm eff} = \int \dd^4 x \, P(X) \; , \qquad X \equiv (\di \phi)^2 \; ,
\ee
where $P$ is a generic function associated with the equation of state. It is the pressure as a function of the (squared) chemical potential. (We refer the reader to refs.~\cite{Son:2002zn, Nicolis:2011pv, Nicolis:2013lma} for further details about this framework.)

\vskip4pt
The simplest weakly coupled UV completion of the superfluid effective theory is the theory of a complex scalar $\Phi(x)$ with a (linearly realized) $U(1)$ symmetry (see also \cite{Babichev:2018twg, Creminelli:2019kjy}), 
\be \label{U(1)}
S = \int{\rm d}^4x\Big(|\di \Phi|^2 - m^2 |\Phi|^2  - \lambda |\Phi|^4 \Big)\; ,
\ee
where $\phi(x)$ arises as the phase of the complex field: $\Phi(x) = \sfrac1{\sqrt 2} \, \rho(x) \, \e^{i \phi(x)}$.

\vskip4pt
If $m^2$ is positive, the origin is stable and the $U(1)$ symmetry is not spontaneously broken in the absence of a chemical potential, $\mu$, for the $U(1)$ charge. In fact, there is a minimum nonzero value of $\mu$ below which the symmetry is unbroken and the ground state charge density is strictly zero. At tree level, that value is simply $m$. For $\mu > m$, the origin becomes unstable, the $U(1)$ symmetry gets broken, and at the same time the system acquires a charge density. 

\vskip4pt
On the other hand, if $m^2$ is negative the origin is unstable already at $\mu =0$. The $U(1)$ symmetry is thus always spontaneously broken, and any nonzero value of $\mu$, no matter how small, yields a nonzero charge density. 

\vskip4pt
In what follows we will treat both of these cases simultaneously. In each case, once the symmetry is broken, the only light degree of freedom is the Goldstone, the radial mode can be integrated out, and the low-energy dynamics are described by an action of the form~\eqref{P(X)}.

\vskip4pt
A slightly more general UV completion is the theory of $N$ real scalars $\Psi_a(x)$, $a=1, \cdots, N$, with an  internal $SO(N)$ symmetry,
\be \label{SO(N)}
S = \int{\rm d}^4x\left(\frac{1}{2}\big\lvert\di \vec \Psi \big\rvert^2 - \frac{m^2}{2} \big\lvert\vec \Psi \big\rvert^2  - \frac{ \lambda}{4} \big|\vec \Psi\big|^4 \right)\; ,
\ee
of which eq.~\eqref{U(1)} is just the $N=2$ version, with $\Phi = \sfrac1{\sqrt 2}(\Psi_1 + i \Psi_2)$.
In order to get a superfluid state, we should switch on a chemical potential for one of the $SO(N)$ charges. Let's call $J_{ab}$ the $SO(N)$ charge that generates  rotations in the $(ab)$-plane and, without loss of generality, consider turning on a chemical potential for the charge associated to the symmetry generator $J_{12}$. 

\vskip4pt
As in the $U(1)$ case, spontaneous symmetry breaking (SSB) will occur for $\mu > m$ if $m^2$ is positive, and for any value of $\mu$ if $m^2$ is negative. Once the symmetry is broken, we have a gapless Goldstone associated with $J_{12}$ and a number of gapped Goldstones associated with the generators that do not commute with $J_{12}$ \cite{Nicolis:2012vf}. In particular, for each {\em pair} of generators that transform as a doublet under $J_{12}$ we get {\em one} gapped Goldstone with gap $\mu$. Such pairs are simply $(J_{1 b}, J_{2 b})$ with $b>2$, and there are $N-2$ of them. 

\vskip4pt
Overall there is one gapless Goldstone, $N-2$ gapped ones, and one radial mode, making up $N$ degrees of freedom---as many as there are in the original multiplet $\vec \Psi$. At low energies we can integrate out all of the them apart from the gapless Goldstone, and we end up once again with the general framework described at the beginning of this section. It is still an open question whether in a situation like this there is any low-energy remnant of the full $SO(N)$ symmetry once everything but the gapless Goldstone has been integrated out \cite{Cuomo:2020gyl}.

\vskip4pt
In the next section we will perform this integrating-out procedure at one-loop order, as outlined in the Introduction.
As emphasized there, the procedure is only predictive if it can be carried out without introducing new counter-terms beyond those needed to renormalize the UV completions. So, for future reference, we report here the one-loop counter-terms for the theories \eqref{U(1)} and \eqref{SO(N)}, which can be computed via standard relativistic QFT methods (see e.g.,~\cite{Peskin:1995ev}). Working in dimensional regularization (DR) and modified minimal subtraction ($\msbar$), they are 
\be \label{counterterms}
\frac{\delta m^2}{\bar m^2} = -2(N+2)   \, \frac{\bar \lambda}{16\pi^2} \frac{1}{d-4} \; , \qquad \frac{ \delta \lambda}{\bar \lambda} = - 2(N+8) \, \frac{\bar \lambda}{16\pi^2} \frac{1}{d-4} \; ,
\ee
where the barred parameters are the renormalized $\msbar$ ones, and the bare parameters appearing in the Lagrangian are $m^2 = \bar m^2 + \delta m^2$ and $\lambda = \bar \lambda + \delta \lambda$.
The case $N=2$ corresponds to the $U(1)$ theory, eq.~\eqref{U(1)}, and---as usual for $\phi^4$ theories---there is no wave-function renormalization at one loop. Moreover, we are ignoring the renormalization of the cosmological constant, which plays no role in the absence of gravity. 

\vskip4pt
The $\msbar$ scheme is  convenient  for computations, but the renormalized couplings it involves are not particularly physical. In particular, they depend on the value of the (arbitrary) renormalization scale $\bar \mu$. As a nontrivial check of our final result,  we will rewrite  the quantum effective action in terms of physical couplings only, in which case the $\bar \mu$ dependence will drop out. To this end, we will need the relationships between $\msbar$ couplings and more physical ones. For the theories at hand they are
\be \label{pole}
\bar m^2 = m^2_{\rm pole} + 2 (N+2) \dfrac{\lambda}{16\pi^{2}} m^{2} \left( \log\dfrac{\bar{\mu}}{m}+ \dfrac{1}{2}\right) \; , \qquad \bar \lambda = \lambda_{\rm thr} + 2 (N+8) \dfrac{\lambda^{2}}{16\pi^{2}} \left( \log\dfrac{\bar{\mu}}{m}+ \dfrac{1}{3}\right)   \; ,
\ee
where $m^2_{\rm pole}$ is the pole mass in the unbroken phase at vanishing chemical potential (in the case that this is stable), and $\lambda_{\rm thr}$ is the amplitude for two-to-two identical-particle elastic scattering at threshold,
\be
i {\cal M}_{2 \to 2} (s = 4m^2, t=0 , u =0) \equiv  - 6 i  \, \lambda_{\rm thr} \; .
\ee
On the other hand, keeping the $\msbar$ couplings around along with the associated $\bar \mu$ dependence will allow us to RG-improve our result at large chemical potential.

\section{Integrating out at one loop}

We now want to carry out the integrating out and matching procedure outlined in the introduction. The goal is to start from a finite (renormalized) ultraviolet theory and derive the low energy EFT by integrating out the heavy modes at one loop.

\subsection{The $U(1)$ theory}
Let's start with the $U(1)$ case, eq.~\eqref{U(1)}. The more general $SO(N)$ case will be an almost trivial generalization. 
We want to rewrite the path integral in polar coordinates,
\be
\Phi(x) = \frac{1}{\sqrt 2} \rho(x) \e^{i \phi(x)} \; ,
\ee
in terms of which the action reads
\be \label{action polar}
S[\rho, \phi] = \int \dd^4 x \left( \frac{1}{2} (\di \rho)^2 + \frac{1}{2} \rho^2 (\di \phi)^2 -\frac{m^2}{2} \rho^2 - \frac{\lambda}{4} \rho^4 \right) \; ,
\ee
and integrate out $\rho$ at one loop. We are  assuming that the $U(1)$ symmetry is spontaneously broken, so that the path integral is peaked comfortably away from the origin. In this case, we can ignore the subtleties that $\rho$ is constrained to be positive, and that $\phi$ is an angular variable which is periodically identified: $\phi = \phi+2 \pi$.\footnote{See, for instance, ref.~\cite{Kleinert} for a discussion of how to deal with topological constraints like these  in path integrals.}

\vskip4pt
First off, notice that from the path integral measure we get a nontrivial UV-divergent determinant,
\be \label{change}
D \Phi D \Phi^* = D \rho D \phi \det \rho(x) = D \rho D \phi \, \e^{\delta^4(0)  \int \dd^4 x \log \rho(x)} \; .
\ee
The delta function is a pure power-law divergence,
\be
\delta^4(0) = \int \frac{\dd^4 p}{(2\pi)^4} = i  \int \frac{\dd^4 p_E}{(2\pi)^4} \equiv i I_0 \; ,
\ee
which vanishes in dimensional regularization. Eventually we will work in dimensional regularization, so we can drop this determinant. Interestingly---as we will see---even if we were to keep it around, to the order we will be working, all its effects would cancel against other one-loop contributions. 
(Like all path-integral determinants, that in eq.~\eqref{change} is formally of one-loop order.)

\vskip4pt
Second, notice that $\phi$ only enters the action \eqref{action polar} in the combination $X = (\di \phi)^2$. So, when we integrate out $\rho$, the derivative expansion will not be in terms of derivatives of $\phi$, but in terms of derivatives of $X$. This means that configurations with slowly varying $X$, regardless of how large $X$ itself is---that is, regardless of how large the first derivative of $\phi$ is---can be dealt with to {\em zeroth} order in the derivative expansion. 
Operationally, this means that we can compute the effective action for the superfluid state to zeroth order in derivatives of $X$, but to {\em all} orders in $X$, in essentially the same way that one computes a Coleman--Weinberg potential.

\vskip4pt
Finally, as mentioned before, it turns out that integrating out a heavy field beyond tree level is no substitute for responsible EFT matching~\cite{Georgi:1993mps}. 
More precisely: if one were to keep all the non-local structure in the low-energy effective action for the light fields, then integrating out a heavy field would be a consistent procedure---it would correspond to doing the first step in the path integral over all the fields that computes correlation functions involving the light fields only. However,  in general one is interested in a truncated derivative expansion of the low-energy effective theory, and that changes completely the UV behavior of loop integrals, and thus the structure of UV divergences. Hence the need for matching observables directly. 

\vskip4pt
Fortunately, instead of computing all correlation functions and matching them one by one, it is possible to match to all orders in the light fields, but to any fixed order in the derivative expansion. The idea is to use the  quantum effective action. Following \cite{Georgi:1993mps}, and specializing to our case, the rule is: compute the one-$\phi$-irreducible quantum effective action in the full theory, $\Gamma_{\rm full}[\phi]$, at some order in the derivative expansion, and demand that, at the same order in the derivative expansion, the low-energy effective action $S_{\rm EFT}[\phi]$ be such that its 1PI quantum effective action, $\Gamma_{\rm EFT}[\phi]$, matches that computed in the full theory:
\be
\Gamma_{\rm EFT}[\phi] = \Gamma_{\rm full}[\phi] \equiv \Gamma[\phi]\; .
\ee
This is a necessary and sufficient condition for all correlation functions of $\phi$ to match between the full theory and the low-energy effective theory.

\vskip4pt
To compute the one-$\phi$-irreducible quantum effective action in the full theory, we separate $\phi$ into a background plus fluctuations,
\be
\phi(x) \to \phi(x) + \pi(x) \; ,
\ee
and perform the path integral over $\pi$ and the radial mode $\rho$, keeping only the contributions that are 1PI for $\pi$:
\be
\e^{i \Gamma_{\rm full}[\phi]} \equiv \bigg(\int D\rho D\pi \, \e^{i S[\rho, \phi+\pi] }\bigg)_{\rm 1P I \; for \; \pi} \,.
\ee

\vskip4pt
 At tree level, this just amounts to solving the classical equation of motion for $\rho$ and substituting it back into the action:
 \be
 \Gamma^{\rm tree}_{\rm full}[\phi] = S[\rho_0[X], \phi] \; ,
 \ee
where $\rho_0[X]$ is the saddle-point solution for $\rho$, for a given $X = (\di \phi)^2$, which obeys
\be
(\Box + m^2 -X) \rho_0 + \lambda \rho_0^3 = 0  \; .
\ee
When $X$ is a constant (ignoring derivatives of $X$), this has two solutions:
\be
\rho^2_0 = 0  \quad \mbox{and}  \quad \rho^2_0 = \frac{X-m^2}{\lambda} \qquad (X={\rm const}.)\; . \label{saddle}
\ee
The first corresponds to the unbroken phase, and is inconsistent with our assumption of having SSB. Thus, we will not consider it further. The second solution corresponds to SSB, and exists only when
\be
X > m^2  \; ,
\ee
which, for positive $X$, is exactly the condition on $\mu$ stated in Section~\ref{UV completions}.

\vskip4pt
Substituting this solution back into the action~\eqref{action polar}, we get the tree-level $\Gamma[\phi]$ to lowest order in derivatives:
\be \label{tree}
 \Gamma^{\rm tree}_{\rm full}[\phi] =\int \dd^4 x \left( \frac{1}{2} \rho_0^2 \, X -\frac{m^2}{2} \rho^2_0 - \frac{\lambda}{4} \rho_0^4 \right)= \int \dd^4 x \, \frac  {\big(X-m^2 \big)^2}{4\lambda} \; ,
\ee
in agreement with \cite{Babichev:2018twg, Creminelli:2019kjy}.

\vskip4pt
At one loop, we have to perturb about the saddle point, $\rho = \rho_0 + h$, expand the action~\eqref{action polar} keeping only the quadratic terms in $h$ and $\pi$,
\be \label{S2}
S_2[h, \pi] = \int \dd^4 x \left( \frac{1}{2}(\di h)^2 + \frac{1}{2} \rho_0^2 \, (\di \pi)^2 + 2 V^\mu \, h \di_\mu \pi -\frac{1}{2} m^2_{\rm eff} h^2 \right)  \; ,
\ee
where we have defined
\be
m^2_{\rm eff} \equiv m^2 - X + 3 \lambda \rho_0^2    \; , \qquad V_\mu \equiv \rho_0 \di_\mu \phi  \, \label{meff} \; .
\ee
We then integrate over $h$ and $\pi$ to obtain the effective action
\be
\e^{i \Delta \Gamma^{\rm 1 \, loop}_{\rm full}[\phi]} = \int D h D\pi \, \e^{i S_2[h, \pi] } \; .
\label{eq:effaction}
\ee
To be clear about the notation: $\rho_0$, $m^2_{\rm eff}$, and $V^\mu$ are all functionals of the background field $\phi(x)$. In the limit where $\di_\mu \phi$ is constant, they become local {functions} of $\di_\mu \phi$ itself, which then are also constant. In particular:
\be \label{meff const}
m^2_{\rm eff} \xrightarrow{X=\,{\rm const.}} 2(X - m^2) \,.
\ee 
From now on, we will take this limit, so that $\rho_0$, $m^2_{\rm eff}$, and $V^\mu$ are all constant.

\vskip4pt
To compute the path integral~\eqref{eq:effaction}, we first change variables for $\pi$ to canonically normalize it
\be
\pi = \varphi/\rho_0 \; , \qquad D \pi = D \varphi \big( \det \rho_0(x)\big)^{-1}\,.
\ee
This exactly cancels the determinant we got from the change of variables in~\eqref{change}, evaluated at the saddle point.\footnote{In fact, had we kept that determinant around, it would have affected the position of the saddle point by an order $\lambda$ shift. However, had we plugged the new saddle point into the action, by stationarity of the action at the {\em unperturbed} saddle point, the value of the action would not have changed at order $\lambda$. So, overall, at this order, all the effects of these two determinants cancel exactly.}
Next, we complete the square for $h$:
\begin{align}
\int D h D\varphi \, \e^{i S_2 } & =  \int D h' D\varphi \, \exp \left[\frac{i}{2}\left( -h' (\Box+m^2_{\rm eff})h' -\varphi \left[\Box + 4\di_\mu \phi \di_\nu \phi\frac{\di^\mu\di^\nu}{\Box+m^2_{\rm eff}} \right] \varphi \right) \right]\, ,
\end{align}
where we have defined $h' \equiv h - 2\partial^\mu\phi \frac{1}{\square+ m^2_{\rm eff} }\partial_\mu\varphi$.
Up to background-independent factors, we therefore have
\be
\begin{aligned}
\e^{i \Delta \Gamma^{\rm 1 \, loop}_{\rm full}[\phi]} & = \det \big(\Box+m^2_{\rm eff} \big)^{-1/2} \times \det \left( \Box + 4 \di_\mu \phi \di_\nu \phi \, \frac{\di^\mu \di^\nu}{\Box+m^2_{\rm eff}} \right)^{-1/2} \\
& = \det \big(\Box \big)^{-1/2} \times \det \left(\Box+m^2_{\rm eff} + 4 \di_\mu \phi \di_\nu \phi \, \frac{\di^\mu \di^\nu}{\Box} \right)^{-1/2}\,.
\end{aligned}
\ee
The first determinant is background independent, so we can drop it. Using standard functional methods \cite{Weinberg:1996kr}, the second determinant yields
\begin{align} \label{main integral}
i \Delta \Gamma^{\rm 1 \, loop}_{\rm full}[\phi]  =  -\frac12 \int \dd^4 x \int \frac{\dd^4 p }{(2\pi)^4} \log{\left(p^2-m^2_{\rm eff} - 4\di_\mu \phi \di_\nu \phi \, \frac{p^\mu p^\nu}{p^2} \right)} \; ,
\end{align}
where the argument of the log is made dimensionless by an arbitrary scale, whose value only affects the cosmological constant.

\vskip4pt
The anisotropic $\di \phi \di \phi$ term inside the log makes computing the momentum integral slightly complicated. We describe the calculation in Appendix \ref{DR integrals}. Working for simplicity in dimensional regularization, as $d \to 4$ the result  is
\be
\begin{aligned}
\Delta \Gamma^{\rm 1 \, loop}_{\rm full}[\phi]   = &\int \dd^4 x \left( -\frac{1}{32\pi^2 (d-4)}  \big( m_{\rm eff}^4 + 2 m_{\rm eff}^2 X + 2 X^2 \big)\right.  \\
&\hspace{.4cm} + \frac{1}{384 \pi^2 } \bigg[  9 m_{\rm eff}^4 +18 m_{\rm eff}^2 X +10 X^2 - 6(m_{\rm eff}^4 + 2 m_{\rm eff}^2 X +2 X^2) \log\left(m_{\rm eff}^2/\bar \mu^2\right)    \\
&\hspace{2.2cm} \left.- 10 \frac{X^3}{m_{\rm eff}^2} \, f\big( -{4X}/{m_{\rm eff}^2} \big)\bigg] \right) 
\; ,
\end{aligned}
\label{one loop}
\ee
where $m^2_{\rm eff}$ depends on $X$ as in eq.~\eqref{meff const}, $\bar\mu$ is the $\msbar$  renormalization scale, and $f$ is a generalized hypergeometric function, which, if needed, can be rewritten in terms of rational functions, square roots, and logarithms:
\be
\begin{aligned}
f(z) & \equiv  \, _3 F_2\left[\left.\begin{array}{c}
1,~~1,~~\sfrac{7}{2}\\
4,~~5
\end{array}\right\lvert z\right] \label{f(z)} \\
 & = \frac{16}5 \left[ \frac{(3 z^2-10 z - 8)(1-\sqrt{1-z} \,)}{z^4} + \frac{z^2+4}{z^3} - 3 \, \frac{z^2 -4z + 8 }{z^3} \log{\bigg(\frac{1+\sqrt{1-z}}{2}} \, \bigg) \right]\; .
\end{aligned}
\ee
In \eqref{one loop}, we have collected all the divergent terms in the first line. We will discuss renormalization of this effective action in Sec.~\ref{renormalization}.

\subsection{The $SO(N)$ theory}
We now consider the $SO(N)$ version of the story, described by the action eq.~\eqref{SO(N)}. It is immediate to generalize the computation we just performed to this case. 
In fact, if one resists the temptation to express the path integral in spherical coordinates, and instead  uses the higher dimensional version of {\em cylindrical} coordinates, the computation is almost the same as in the $U(1)$ case, with one minor addition. 

\vskip4pt
Consider rewriting the field multiplet $\vec \Psi$ in polar coordinates in the (12)-plane, but in cartesian coordinates for the other components:
\be
\vec \Psi(x) =  \left( \begin{array}{c}
\rho(x) \, \cos \phi(x) \\ \rho(x) \, \sin \phi(x) \\ \sigma_1(x) \\ \vdots \\ \sigma_{N-2}(x)
\end{array}\right) \,.
\ee
Like in the $U(1)$ case, the path integral measure changes as
\be \label{change SO(N)}
D \vec \Psi \equiv D \Psi_1 \dots D \Psi_N  = D \rho \, D \phi \, \det \rho(x)  \times D \vec \sigma \,  \; ,
\ee
where $\vec \sigma$ is the $(N-2)$-dimensional vector $(\sigma_1, \dots, \sigma_{N-2})$,
and the action becomes
\be
S[\rho, \phi, \vec \sigma \, ] = \int \dd^4 x \left(\frac{1}{2}(\di \rho)^2 + \frac{1}{2} \rho^2 (\di \phi)^2 + \frac{1}{2}(\di \vec \sigma)^2-\frac{ m^2}{2} \big(\rho^2 + |\vec \sigma|^2\big) - \frac{\lambda}{4} \big(\rho^2 + |\vec \sigma|^2\big)^2 \right) \; .
\ee
Notice that $\vec \sigma$ does not couple directly to $\phi$. This is the main advantage of using this parametrization of the field space.

\vskip4pt
The change in measure can be neglected for exactly the same reason as before---it vanishes in dimensional regularization, and more generally, it cancels against another determinant that we will encounter along the way. To compute the quantum effective action for $\phi$, we therefore have to evaluate the path integral
\be
\e^{i \Gamma_{\rm full}[\phi]} \equiv \left(\int D\rho \,D\pi\, D \vec \sigma \, \e^{i S[\rho, \, \phi+\pi, \, \vec \sigma \,] }\right)_{\rm 1P I \; for \; \pi} \,.
\ee
Working again to zeroth order in derivatives of $X = (\di \phi)^2$, the saddle point for $\rho$ is exactly the same as before, eq.~\eqref{saddle}, and for $\vec \sigma$ it is just $\vec \sigma=0$. This means that the tree-level $\Gamma[\phi]$ is identical to before, eq.~\eqref{tree}.

\vskip4pt
We perturb the action around this saddle point as $\rho = \rho_0 + h$ and $\phi = \phi+\pi$ and expand up to quadratic order. After doing this, the action becomes
\be
\label{eq:sonperturbed}
S_2[h, \pi, \vec \sigma \, ] =  \int \dd^4 x \left(  \frac{1}{2} (\di h)^2 + \frac{1}{2}  \rho_0^2 \, (\di \pi)^2 + 2 V^\mu \, h \di_\mu \pi -\frac{1}{2} m^2_{\rm eff} h^2   + \frac{1}{2}(\di \vec \sigma )^2 - \frac{1}{2}X  |\vec \sigma|^2 \right) \; .
\ee
with $V^\mu$ and $m_{\rm eff}^2$ defined in the same way as in eq.~\eqref{meff}. 
It is now clear that, at this order, we have two decoupled sectors:
\begin{itemize}
\item
{\bf  The $\boldsymbol \pi$-$\boldsymbol h$ sector:} This sector is given by the first four terms of~\eqref{eq:sonperturbed} and has precisely the same action as in the $U(1)$ case, eq.~\eqref{S2}. Correspondingly, the path integral over these fields yields exactly the same result as in the $U(1)$ case, eq.~\eqref{one loop}.
\item
{\bf The $\boldsymbol{\vec \sigma}$ sector:} This sector is given by the last two terms of~\eqref{eq:sonperturbed} and is made up of $N-2$ free massive scalars with mass $X$---the gapped Goldstones---in agreement with the general results  of \cite{Nicolis:2011pv}, which we summarized briefly in Sec.~\ref{UV completions}. Up to irrelevant field-independent factors, the path integral over this sector yields the determinant
\be
\det(\Box + X)^{-\frac{N-2}2}  =  \exp\left( -\frac{N-2}2  \int \dd^4 x \int \frac{\dd^4 p }{(2\pi)^4} \log{\big(p^2- X\big)} \right) \; ,
\ee
which  corresponds to one of the momentum integrals we have already computed for the $U(1)$ case (see~\eqref{eq:int2} in Appendix~\ref{DR integrals}). The contribution from the log is then 
\begin{align}
 \int \frac{\dd^4 p_E }{(2\pi)^4} \log{\big(p_E^2+X \big)} = \frac{X^2}{16\pi^2} \left( 
\frac{1}{d-4} - \frac{3}{4 }    + \frac{1}{2}  \log(X/\bar \mu^2) \right) \; .
\end{align}
\end{itemize}
Putting everything together, the quantum effective action for $\phi$ in the $SO(N)$ case simply reads
\be \label{Gamma SO(N)}
\Gamma_{\rm full}^{SO(N)}[\phi] = \Gamma_{\rm full}^{U(1)}[\phi] - (N-2)\int \dd^4 x \, \frac{X^2}{32\pi^2} \left( 
\frac{1}{d-4} - \frac{3}{4 }    + \frac{1}{2}  \log(X/\bar \mu^2) \right) \; ,
\ee
where $\Gamma_{\rm full}^{U(1)}[\phi]$ is the sum of the tree level effective action, eq.~\eqref{tree}, and of the one-loop correction in the $U(1)$ case, eq.~\eqref{one loop}.

\subsection{Renormalization}\label{renormalization}

We now come to the issue of renormalization. We will consider directly the more general $SO(N)$ case. The $U(1)$ case can be obtained by simply setting $N=2$.

\vskip4pt
The divergent terms in \eqref{Gamma SO(N)}---including those that come from \eqref{one loop}---read
\be
\begin{aligned}
\Gamma_{\rm full}^{SO(N)}[\phi] & \supset -\frac{1}{32\pi^2 (d-4)}  \big( m_{\rm eff}^4 + 2 m_{\rm eff}^2 X + N X^2 \big)\,,  \\
& \xrightarrow{m_{\rm eff}^2\to\, 2(X - m^2)} \frac{1}{32\pi^2 (d-4)}  \big(12  m^2 X - (N+8) X^2 \big) \label{divergences}  \; ,
\end{aligned}
\ee
where in the last step we dropped an $X$-independent term, since it is associated with the cosmological constant, which we haven't been keeping track of.

\vskip4pt
In the $\msbar$ scheme, one would normally just drop these divergent terms. However,  as emphasized in the Introduction, we do not want to introduce new counterterms besides those already needed to renormalize the UV completion, eq.~\eqref{SO(N)}, which  is defined in terms of two parameters only, $m^2$ and $\lambda$.\footnote{In principle, one should also consider wave-function renormalization, but in $\phi^4$ theories this starts mattering only at two-loop order.} 
So, it had better be the case that renormalizing $m^2$ and $\lambda$ as in eq.~\eqref{counterterms} completely cancels the divergent terms in~\eqref{divergences}. 

\vskip4pt
To check this, rewrite the bare couplings in the tree-level effective action \eqref{tree} as
\be
m^2 = \bar m^2 + \delta m^2 \; , \qquad\qquad \lambda = \bar \lambda + \delta \lambda \; ,
\ee 
where $\bar m^2$ and $\bar \lambda$ are the $\msbar$ renormalized couplings (the relevant counterterms are given by~\eqref{counterterms}). Then,  expand eq.~\eqref{tree} to first order in $\delta m^2$ and $\delta \lambda$. It is a matter of simple algebra to check that the only effect of this procedure is to cancel the divergent terms \eqref{divergences}, apart again from a cosmological constant-related, $X$-independent term.

\vskip4pt
In conclusion, the quantum effective action for $\phi$ to one-loop order and to zeroth order in derivatives of $X$ is simply eq.~\eqref{Gamma SO(N)}, with the divergent terms removed and with all the couplings (included the $m^2$ inside $m^2_{\rm eff}$) now interpreted as renormalized $\msbar$ couplings. Admittedly, its explicit form---which we display in the next subsection---is not particularly illuminating. Before discussing different limits and special cases to extract some physical insights, we briefly address the issue of matching to the low-energy effective theory---an issue which, at this order,  happens to be trivial in dimensional regularization.

\subsection{Matching}
The matching prescription suggested by~\cite{Georgi:1993mps}, which we briefly reviewed in the Introduction, instructs us to engineer a low-energy effective action for $\phi$ that---at one loop and to zeroth-order in derivatives of $X$---yields the same quantum effective action as the one we derived above starting from the UV completion. 

\vskip4pt
Let's start then with the most general lowest-order effective action for $\phi$, eq.~\eqref{P(X)}, and define the one-loop quantum effective action associated with it:
\be
\e^{i \Gamma[\phi]}  = \bigg(\int D\pi \, \e^{i S_{\rm eff}[\phi+\pi] }\bigg)_{\rm 1P I} = \e^{i S_{\rm eff}[\phi]} \int D\pi \, \e^{i S_{2}[\pi]}  \; ,
\ee
where $S_2$ is the quadratic action for the fluctuations of $\phi$:
\be
S_2[\pi] = \int \dd^4 x \, \frac{1}{2} Z^{\mu\nu} \, \di_\mu \pi \di_\nu \pi \;  \; , \quad{\rm with}\quad Z^{\mu\nu} \equiv 2 P'(X) \eta^{\mu\nu} + 4 P''(X)\di^\mu \phi \di^\nu \phi \; .
\ee
We thus get (up to field-independent factors)
\begin{align}
\int D\pi \, \e^{i S_{2}[\pi]} & = \det(Z^{\mu\nu} \di_\mu \di_\nu )^{-1/2} \\
& = \exp\left(- \frac12 \int \dd^4 x \int \dd^4 p \, \log \big(Z^{\mu\nu} p_\mu p_\nu\big) \right) \; ,
\end{align}
where in the last step we have used that, for constant $\di_\mu \phi$, the matrix $Z^{\mu\nu}$ is constant.

\vskip4pt
The momentum integral above vanishes in dimensional regularization, simply because it's a scale-less integral. Explicitly,
denoting the integrand by $I(p)$ and using standard properties of dimensional regularization \cite{Collins:1984xc}, we have that a rescaling of the integration variable yields
\be
\int \dd^d p \, I(p) = \alpha^d \int \dd^d p \,  I (\alpha p) = \alpha^d \int \dd^d p \,  \big(I (p) + 2 \log \alpha \big) =  \alpha^d \int \dd^d p \,  I (p) \;,
\ee
which implies that the integral of $I(p)$ must vanish.

\vskip4pt
This implies that, if one uses dimensional regularization, at one-loop order and to zeroth order in derivatives of $X$, the quantum effective action is the same as the tree-level low-energy action:\footnote{This is also consistent with the analyses of~\cite{Nicolis:2004qq, deRham:2014wfa,Goon:2016ihr} which show that, in dimensional regularization, the first loop corrections scale schematically as ${\cal L} \sim \partial^4 X^n$.
}
\be
\Gamma[\phi] = S_{\rm eff}[\phi] \; \qquad \qquad \mbox{(in dim. reg.)} \; .
\ee
Reading this equation backwards, we see that, to this order, the low-energy effective action we should use for $\phi$ is simply the $\Gamma[\phi]$ we have computed in the previous subsections. 

\vskip4pt
In conclusion, integrating out at one loop all the heavy fields in the $SO(N)$ theory \eqref{SO(N)} yields the Goldstone effective theory
\begin{align} 
\nonumber
P(X) &=   \frac  {\big(X-m^2 \big)^2}{4\lambda} \\\nonumber
&\hspace{.5cm}+ \,\frac{1}{384 \pi^2 } \bigg[  -8 X^2 + (m_{\rm eff}^4 + 2 m_{\rm eff}^2 X +2 X^2) \big(9-6 \log(m_{\rm eff}^2/\bar \mu^2)   \big) - 10 \frac{X^3}{m_{\rm eff}^2} \, f\left( -\frac{4X}{m_{\rm eff}^2} \right)  \\
&\hspace{2.5cm}+ (N-2) X^2 \big(9-6  \log(X/\bar \mu^2) \big) \bigg]  \; ,
\label{final P(X)}
\end{align}
where $m^2_{\rm eff} = 2(X-m^2)$, $f$ is given in eq.~\eqref{f(z)}, and $m^2$ and $\lambda$ are to be interpreted as the renormalized $\msbar$ couplings of theory \eqref{SO(N)}, evaluated at $\bar \mu$.
The first line is the tree-level contribution. The second line is the one-loop correction in the $U(1)$ case. The third line is the only term that, at one-loop, differentiates the $SO(N)$ case from the $U(1)$ one.

\section{Some physics}

The final result for the Goldstone effective action, eq.~\eqref{final P(X)}, is not particularly illuminating in and of itself.  We can however explore some interesting physical questions and  different regimes using it. These explorations will also serve as consistency checks.

\subsection{$P(X)$ in terms of physical parameters}

For constant $X=\mu^{2}$, the function $P(X)$ is associated with the zero-temperature equation of state of the relativistic superfluid, and gives the pressure as a function of the chemical potential. Since this is an observable quantity, the apparent dependence on the renormalization scale $\bar \mu$ must disappear once we express the final result \eqref{final P(X)} in terms of physical quantities.

\vskip4pt
We present the result using physical parameters in the unbroken phase at zero chemical potential, which are meaningful for theories with positive $m^{2}$.
We do so by trading the (unphysical) $\msbar$ mass and coupling parameters for the pole mass and the two-to-two scattering amplitude at threshold, as in equation~\eqref{pole}. 

\vskip4pt
After straightforward manipulations, we obtain:
\be
\label{physical_PX}
\begin{aligned}
P(X) =  \; & \frac  {\big(X-m^2_{\rm pole} \big)^2}{4\lambda_{\rm thr}} 
+ \,\frac{1}{384 \pi^2 } \Bigg[-8 (6X^{2}-4m^2_{\rm pole}X-m^4_{\rm pole}) \\[1pt]
&\hspace{1.25cm}+  \,2 (5 X^{2} - 6 m^2_{\rm pole} X + 2 m^4_{\rm pole}) \big(9-6 \log(m_{\rm eff}^2/m^2_{\rm pole})   \big) 
- 10 \frac{X^3}{m_{\rm eff}^2} \, f\left( -\frac{4X}{m_{\rm eff}^2} \right) \\[1pt]
&\hspace{1.25cm}- (N-2)\Big(\,4  (X-m^2_{\rm pole})(X+2m^2_{\rm pole}) - X^2\left[9-6 \log(X/m^2_{\rm pole})\right]\Big) \Bigg]  \; ,
\end{aligned}
\ee
where we have discarded a constant term that can be reabsorbed into the cosmological constant.\footnote{From now on, all the expressions for $P(X)$ are understood to be given up to a cosmological constant term.} As expected, all the $\bar \mu$ dependent terms of equation~\eqref{final P(X)} cancel in a nontrivial way, confirming the consistency of our result.

\subsection{SSB for positive $m^2$}\label{SSB section}
As reviewed in Sec.~\ref{UV completions}, our theories exhibit SSB at the classical level either for negative $m^2$, or, if $m^2$ is positive, when the chemical potential exceeds $m$.  This matches the fact that for relativistic free bosons at zero temperature, Bose--Einstein condensation (BEC) happens precisely at $\mu = m$.

\vskip4pt
How do these statements get modified at one-loop? In a forthcoming paper, we will study in detail the relationship between finite charge density and SSB for scalar theories at the quantum level. 
For the time being, we content ourselves with showing that, in cases with positive $m^2$, the one-loop Goldstone effective theory is well-behaved if and only if $X > m^2_{\rm pole}$, where $m_{\rm pole}$ is the one-loop pole mass. Since $X$ can be thought of as the local value of the (squared) chemical potential---see Sec.~\ref{UV completions}---and since a healthy Goldstone effective theory can be taken as a sign of having SSB, this suggests that at the quantum level the criterion for having SSB in scalar theories is that the chemical potential exceeds the physical mass of the bosons at hand. Which, although not obvious, is at least reassuring from an intuitive BEC perspective.\footnote{We refer the reader to our forthcoming paper for further discussions about this point.}

\vskip4pt
To begin with, let's see how we can recover the fact that we have SSB only for $X > m^2$ from the tree-level result,
\be
P_{\rm tree} (X)= \frac{(X-m^2)^2}{4\lambda} \; .\label{Ptree}
\ee
The idea is to check the stability of a configuration with constant $X = \mu^2$. We  separate the Goldstone field into  background plus fluctuations as
\be
\phi(x) = \mu t + \pi(x) \; ,
\ee
and then expand the effective action to quadratic order in the perturbations $\pi(x)$. For a generic $P(X)$, we get
\be
P(X) \supset \left[P'(\mu^2) + 2 P''(\mu^2) \mu^2 \right] \dot \pi^2 -  P'(\mu^2) (\vec \nabla \pi)^2 \equiv \alpha\, \dot \pi^2 - \beta\, (\vec \nabla \pi)^2 \; .
\ee
Stability corresponds to having both $\alpha$ and $\beta$ be positive. The sound speed is $c_s^2 = \beta/\alpha$, and so subluminality corresponds to $\alpha \ge \beta$.
Overall, the combined conditions to have both stability and subluminality thus read
\be \label{conditions}
P'(\mu^2) > 0  \quad \mbox{and} \quad  P''(\mu^2) \ge 0 \; .
\ee 
For the tree level effective action \eqref{Ptree}, we have
\be
P_{\rm tree}'(\mu^2) = \frac{(\mu^2-m^2)}{2\lambda} \ ; \qquad P_{\rm tree}''(\mu^2) = \frac{1}{2\lambda} \; . 
\ee
From this, we see that second condition in \eqref{conditions} ($P''>0$) is always obeyed, but the first condition (that $P'>0$) is obeyed only for $\mu^2 > m^2$, as expected.

\vskip4pt
Let's now check what happens at one-loop. Since in perturbation theory the one-loop effective action is close to the tree-level one, we can restrict to the neighborhood of $X = m^2$, and see what the stability and subluminality conditions look like in that neighborhood. Expanding eq.~\eqref{final P(X)}  to quadratic order in $(X-m^2)$, choosing for simplicity the renormalization scale as $\bar \mu = m$, we get
\be \label{NR}
P_{\rm 1-loop}(X) \simeq \frac{1}{\lambda} \left[ \frac{\lambda}{32\pi^2} (N+2) m^2 (X-m^2) + \frac14  (X-m^2)^2 \left( 1 + \frac{\lambda}{4\pi^2}\right) \right] \; ,\qquad(\bar\mu^2 = m^2) \; .
\ee
From this we can compute the first and second derivatives:
\begin{align}
P'_{\rm 1-loop}(X) & \simeq \frac{1}{\lambda} \left[ \frac{\lambda}{32\pi^2} (N+2) m^2  + \frac12  (X-m^2) \left( 1 + \frac{\lambda}{4\pi^2}\right) \right]\,,  \\
P''_{\rm 1-loop}(X) & \simeq \frac{1}{2\lambda} \left( 1 + \frac{\lambda}{4\pi^2}\right) \; .
\end{align}
From this we see that, 
once again, the second condition in~\eqref{conditions} is always obeyed, but now $P'>0$ is obeyed only for (to first order in $\lambda$)
\be
X > m^2 \left(1- \frac{\lambda}{16\pi^2} (N+2)\right) \; ,
\ee 
which is precisely the pole mass, eq.~\eqref{pole}, for $\bar \mu = m$.\footnote{Recall that, compared to eq.~\eqref{pole}, we are now using a lighter notation whereby $m$ and $\lambda$  stand for the $\msbar$ couplings $\bar m$ and $\bar \lambda$ evaluated at scale $\bar \mu$.} 

\vskip4pt
The same result can be recovered working directly in terms of the physical parameters. Using the $P(X)$ of eq.~\eqref{physical_PX}, and expanding around $X=m^{2}_{\rm pole}$ we obtain equation~\eqref{eq:PX_positive_onshell} below. Performing a stability analysis along the same lines, we find that $P'(\mu^{2})>0$ if and only if $X> m^{2}_{\rm pole}$. 

\subsection{Low densities}\label{low densities}
An interesting limit to consider is that of low charge densities. In this limit, the two cases $m^2  < 0$ and $m^2 > 0$ behave in strikingly different ways:
\begin{itemize}
\item {\bf Negative $\boldsymbol{m^2}$:} In this case we have SSB already for $X=0$, which corresponds to a Poincar\'e invariant ground state with vanishing charge density. For small $X \ll -m^2$, one gets a small charge density of order $X$. Notice that this case just corresponds to a Lorentz-invariant theory of Goldstones, expanded about a small Lorentz-violating background. The radial mode that has been integrated out had a mass $m^2_{\rm eff} \simeq -2 m^2$, and so this is the scale that controls the derivative expansion in the $U(1)$ case. For the more general $SO(N)$ case, if one insists on integrating out the gapped Goldstones as well---like we have done---the regime of validity of the effective theory is much narrower, since the derivative expansion is controlled by the gap of the gapped Goldstones, which is $X \ll -m^2$. Related to this, the effective theory \eqref{final P(X)} for the gapless Goldstone $\phi$  is not analytic at $X=0$,
\be
\begin{aligned}
P(X) \simeq \frac{1}{\lambda} \bigg[ &- \frac12 m^2 X \left( 1 + 7 \frac{\lambda}{16\pi^2}  \right) + \frac1{4}X^2 
\left( 1 + \frac{11}{4}\frac{\lambda}{16\pi^2} \right)\\
& \hspace{.5cm}~~+ \left(\dfrac{N-2}{4}\right) \dfrac{\lambda}{16\pi^{2}} X^2 
\left( \dfrac{3}{2} - \log \left( -\dfrac{X}{2 m^{2}}\right) \right)
 \bigg]\,,  \qquad\quad (\bar\mu^2 = -2m^2) \; ,
\end{aligned}
\ee
where for simplicity we set $\bar\mu^2 = -2m^2$. The non-analyticity $\sim X^2 \log X$ is a sign that we have integrated out degrees of freedom that become gapless as $X \to 0$.

\item {\bf Positive $\boldsymbol{m^2}$:} As we saw in Sec.~\ref{SSB section}, in this case we have SSB and an associated charge density only for $X > m^2_{\rm pole}$. It is convenient in this case to express the $P(X)$ in terms of the on-shell parameters. Close to that lower limit we have
\be
\label{eq:PX_positive_onshell}
P(X) =\left(\dfrac{1}{4\lambda_{\rm thr}} - \dfrac{N+2}{96\pi^{2}}\right) (X-m_{\rm pole}^2)^2  - \dfrac{\sqrt{2}}{15 \pi^{2}} \dfrac{(X-m^{2}_{\rm pole})^{5/2}}{m_{\rm pole}} + \mathcal{O}\left((X-m^{2}_{\rm pole} )^{3}\right)\; .
\ee
Now the charge density is of order $(X-m^2_{\rm pole})$. However, in contrast to the negative $m^2$ case, here the  Goldstone effective theory is highly non-relativistic at low densities. In particular, the speed of sound,
\be
c_s^2 = \frac{P'(X)}{P'(X) + 2P''(X)X} \; ,
\label{eq:speedofsound}
\ee
is found to be
\begin{equation}
c_{s}^{2} \simeq \dfrac{X-m_{\rm pole}^2}{3X-m_{\rm pole}^2} \simeq \dfrac{X-m_{\rm pole}^2}{2 m_{\rm pole}^2},
\end{equation} 
which goes to zero when the density goes to zero, while in the negative $m^2$ case it goes to unity in that limit, since there one recovers Lorentz invariance. 

The structure of the derivative expansion is also different: here the gap of the radial mode is $m^2_{\rm eff} \simeq 2(X- m^2)$, which is small, while the gap of the gapped Goldstones is $X \simeq m^2$, which is not small. 
The leading non-analytic effect is thus associated with the radial mode, which has been integrated out, and is the branch cut of $(X-m^{2}_{\rm pole})^{5/2}$. Notice that the structure of non-analyticities is different from the negative $m^2$ case: the $\log (m^2_{\rm eff})$ in eq.~\eqref{physical_PX} is cancelled exactly by another one that emerges in the small-$m^2_{\rm eff}$ expansion of $f(-4X/m^{2}_{\rm eff})$---see eq.~\eqref{f(z)}. Such a perfect cancellation is quite mysterious, and we are not sure what its physical interpretation is.

\end{itemize}

\subsection{High densities: RG improvement}

An advantage of having the effective action to all orders in $X$ is that we can push it to very large values of $X$---or equivalently very high densities---where things drastically simplify. In fact, the derivative expansion becomes better and better in that limit, since the radial mode (with gap $m^2_{\rm eff} \simeq 2 X$) and the gapped Goldstones (with gap $X$) become very heavy. From \eqref{final P(X)}, we find that the leading terms in this limit are
\be \label{large X}
P(X) \simeq \frac{1}{4 \lambda} X^2 \left[ 1 + \frac{\lambda}{96 \pi^2} \Big(82 -5 \, \kappa -60 \log(2X/\bar \mu^2 ) + (N-2)\left[9-6\log(X/\bar \mu^2 ) \right] \Big)\right]  \; ,
\ee
where $\kappa= {}_3F_2\left[\left.\begin{array}{c}
1,~~1,~~\sfrac{7}{2}\\
4,~~5
\end{array}\right\lvert -2\right]=0.771885\dots$ is a constant.

\vskip4pt
Notice that, given the one-loop renormalization group (RG) equation for $\lambda$,
\be
\bar \mu \frac{\dd \lambda}{\dd \bar \mu} = \Big(20+2(N-2)\Big) \frac{\lambda^2}{16 \pi^2} \,,
\ee
combined with the fact that
 there is no wave-function renormalization for $\phi$, we have 
\be
\bar \mu \frac{\dd}{\dd\bar \mu} P(X)  = 0 \; ,
\ee
as expected. Then, when the log in  \eqref{large X} becomes large, we can use the RG and resum the leading logs to obtain
\be
\label{eq:highdensity}
P(X) \simeq \frac{X^2}{4 \, \lambda(2X)} \; ,
\ee
where $\lambda(2X)$ is the running coupling evaluated at scale $\bar{\mu}^{2}=2X$. This is the same as the leading high-$X$ behavior of the tree-level effective action, \eqref{Ptree}, but with the coupling constant replaced by its running value at the appropriate scale.

\vskip4pt
It is interesting to compute the speed of sound~\eqref{eq:speedofsound}.
To this order, we get
\be
\label{eq:cs2}
c_s^2 \simeq \frac13 + \frac19 \beta_1 \, \lambda(2X) \; ,
\ee
where $\beta_1$ is the one-loop $\beta$-function coefficient, $\beta_1 = \frac{20+2(N-2)}{16 \pi^2}$. Notice that we are slightly above the famous value of $1/3$ for a conformally invariant fluid, and more and more so as we run towards the UV, since $\lambda$ grows. 
This behavior can be understood by noticing that, for constant $\lambda$, the effective lagrangian~\eqref{eq:highdensity} yields a scale invariant theory if one assigns $\phi(x)$ scaling dimension zero. Indeed, it corresponds to the effective action of a conformal superfluid in $3+1$ dimensions~\cite{Hellerman:2015nra,Monin:2016jmo}. The system is thus conformally invariant, up to scaling-violating effects induced by the running of the couplings. This allows us to relate the corrections to the sound speed to the trace anomaly.

\vskip4pt
The energy momentum tensor of a general superfluid is given by
\begin{equation}
T_{\mu\nu} = 2 P'(X) \partial_{\mu}\phi \, \partial_{\nu}\phi - \eta_{\mu\nu} P(X) \; , 
\end{equation}
from which we can read off its energy density and pressure as
\begin{equation}
\varrho = 2P'(X) X -P(X) , \qquad \qquad p=P(X).
\end{equation}
For the effective action of eq.~\eqref{eq:highdensity}, the trace of $T_{\mu\nu}$ reads
\begin{equation}
\label{eq:traceanomaly}
T^{\mu} {}_{\mu}= -\beta_{1} \lambda(2X) P(X) \; ,
\end{equation}
so that the system has a trace anomaly proportional to the $\beta$-function of $\lambda$. On the other hand, the trace of the energy momentum tensor can be expressed in terms of the energy density and pressure as $T^{\mu} {}_{\mu}=\varrho-3p$, so that its $X$ derivative is directly related to $c_s^2$'s departure from $1/3$:\footnote{For fluids and superfluids alike, the sound speed is related to the energy density and the pressure by $c_{s}^{2}= \frac{{\rm d}p}{{\rm d}\varrho}$.}
\begin{equation}
\frac{\dd T^{\mu} {}_{\mu}}{\dd X} = -3 \frac{\dd \varrho}{\dd X} \, \left(c_s^2 - \frac{1}{3}\right) = 3  \frac{\dd p}{\dd X} \left( \dfrac{1}{3c_s^2} -1 \right).
\end{equation}
Equating this to the $X$ derivative of eq.~\eqref{eq:traceanomaly} and working to lowest order in $\lambda$---and so, in particular, neglecting the $X$ dependence of $\lambda$---we reproduce exactly eq.~\eqref{eq:cs2}.


\subsection{A cross-check}

We can also use some of the results of \cite{Kourkoulou:2021ksw} as a nontrivial check of ours. In particular, one of those results is that for 
a superfluid with $P(X) = \frac12 X$ coupled to a heavy field $h(x)$ as
\be \label{L from framid paper}
S= \int {\rm d}^4 x \left( \frac{1}{2} X  + X h + \frac{1}{2} (\partial h)^2 - \frac{ M^2}{2} h^2 \right)\; , 
\ee
the one loop correction to the expectation value of the stress-energy tensor to first order in the background $X$ obeys\footnote{We are writing eqs.~\eqref{L from framid paper} and \eqref{rho+p from framid paper} not in the original notation of~\cite{Kourkoulou:2021ksw}, but rather in variables that will make the comparison with our results more straightforward.
}
\be \label{rho+p from framid paper}
\Delta(\varrho + p) \simeq - \frac{X M^2}{4 \pi^2 \, d} \bigg( \frac{M^2}{\e^{\gamma_E} \bar \mu^2}\bigg)^\frac{d-4}{2} \Gamma\bigg( \frac{2-d}{2} \bigg) \; ,
\ee
where $\gamma_E$ is the Euler--Mascheroni constant.
Now, the action in the $U(1)$ case, before we integrate out the radial mode, looks very much like \eqref{L from framid paper}. In particular, if we consider the quadratic expansion \eqref{S2} about a non-trivial background $\phi(x)$ and the associated saddle point for $\rho(x)$, and if we canonically normalize $\pi$, we get
\be
S_2[h, \pi] = \int \dd^4 x \left( \frac{1}{2} (\di \pi)^2+ \frac{1}{2} (\di h)^2  + 2 \partial^\mu \phi \, h \di_\mu \pi -\frac{1}{2}m^2_{\rm eff} h^2 \right)  \; .
\ee
This is {\em exactly} the same as the quadratic expansion of \eqref{L from framid paper}, if we formally treat $m^2_{\rm eff}$---which secretly depends on $X$---as a constant: $m^2_{\rm eff} \to M^2$.
In this case, the one-loop correction to the quantum effective action \eqref{one loop}, to first order in the background $X$ and neglecting a cosmological constant, reads
\be \label{linear X}
\Delta \Gamma (X) \simeq \alpha \, M^2  X \; , \quad{\rm with}\quad \alpha \equiv -\frac1{16\pi^2(d-4)}+\frac{3-2 \log\big( M^2/\bar\mu^2 \big)}{64 \pi^2} \; .
\ee
The stress-energy tensor that we get by formally treating this as a classical Lagrangian is nothing but the one-loop correction to the {\em expectation value} of the stress energy tensor, simply because  for a general field theory one has
\be
\langle T_{\mu\nu} (x)\rangle = \frac2{\sqrt{-g}} \frac{\delta \Gamma}{\delta g^{\mu\nu}(x)} \; ,
\ee
as can be checked straightforwardly using standard functional methods.

\vskip4pt
From  \eqref{linear X} we thus get
\be
\Delta(\varrho + p) = 2 \,\Delta \Gamma' (X) \,X \simeq 2 \alpha \, M^2  X \; ,
\ee
which is exactly the $d\to4$ limit of eq.~\eqref{rho+p from framid paper}. Notice that the fact that we consistently discarded contributions to the cosmological constant does not affect our ability to compute the value of $\varrho + p$ precisely because such a combination is insensitive to the cosmological constant, which always has $\varrho + p =0 $ as a consequence of Lorentz invariance

\vskip4pt
The result \eqref{rho+p from framid paper} was derived in \cite{Kourkoulou:2021ksw} with methods that are so different from ours, that we find our being able to reproduce it to be a very non-trivial check of the validity of our analysis (and of that of \cite{Kourkoulou:2021ksw} as well).


\section{What happened to $SO(N)$?}

Consider once again the generic $SO(N)$ case. After renormalization, the low-energy effective theory is given by eq.~\eqref{final P(X)}. Recall that the only difference with the $U(1)$ case is the relatively simple term in the last line, explicitly proportional to $N-2$. This creates a puzzle, or at least an uncomfortable situation, already pointed out in \cite{Cuomo:2020gyl}: what happened to the full $SO(N)$ symmetry? It seems extremely unlikely that adding a simple term like the $(N-2)$ one in the last line of~\eqref{final P(X)} can promote a quite complicated structure like that in the first two lines from being only symmetric under $U(1)$  to being symmetric under the full $SO(N)$---a much bigger, more complicated (i.e., non-abelian) group.  

\vskip4pt
The only sensible answer at the moment seems to be that, in the low-energy effective theory without the gapped Goldstones,  there is simply {\em no} trace of the original $SO(N)$ symmetry. 
To be clear, as we already emphasized in Sec.~\eqref{low densities}, the fact that the last line of \eqref{final P(X)} is not analytic at $X=0$ is a sign that some degrees of freedom that have been integrated out become gapless at $X=0$, and maybe it can be argued that the $N-2$ in front is a sign that there are $N-2$ of them. But it is certainly not clear how one could conclude from these observations that these degrees of freedom must be the gapped Goldstones of a spontaneously broken $SO(N)$ theory, simply because it is not clear in what sense the effective action \eqref{final P(X)}  is invariant under $SO(N)$. 

\vskip4pt
Notice that, with our conventions (see Sec.~\ref{UV completions}), a state with a nontrivial $X$ corresponds to having a finite chemical potential for the $J_{12}$ generator of $SO(N)$. Of the other $SO(N)$ generators, $2 (N-2)$ of them don't commute with $J_{12}$, but the remaining ones do. In fact, the ones that do are {\em not} spontaneously broken, and so all the Goldstones can be taken to be in linear representations of them. In particular, the gapless Goldstone that we retain in the low-energy theory is neutral under them, since it has the same quantum numbers as $J_{12}$, which commutes with those generators. So, for the generators that commute with $J_{12}$ there is no mystery; there is no trace of these symmetries in the effective theory for the gapless Goldstone, because it so happens that they only act nontrivially on gapped fields.
Nothing in the low-energy theory transforms under them. This is the same, for instance, as noticing that there is no trace of baryon number in the pions' chiral Lagrangian, simply because all baryons are heavy and do not appear in the low-energy theory. 

\vskip4pt
On the other hand, the $2 (N-2)$ generators of $SO(N)$ that do not commute with $J_{12}$  {\em do} act non-trivially on the gapless Goldstone, but they do so in a way that depends on the gapped Goldstones. When these are integrated out, it's not clear how to make those generators act on the gapless Goldstone. In fact, there is no known way to realize non-linearly $SO(N)$ with just a single field, even accounting for the subtleties associated with the spontaneous breaking of spacetime symmetries. That is, sometimes the number of Goldstones needed to realize a given symmetry breaking pattern can be reduced by  so-called inverse Higgs constraints, which can be physically  interpreted as the result of integrating out certain gapped Goldstones. But the rules of when that is possible are well understood, and this is not one of those cases (for a review, see for instance~\cite{Nicolis:2013sga}).

\vskip4pt
At least in perturbation theory, the technical reason we lose the non-abelian $SO(N)$ symmetry after integrating out the gapped Goldstones seems to be the following. The gapped Goldstones transform non-linearly under $SO(N)$. So, whatever boundary conditions we choose for them  {\em cannot} be invariant under $SO(N)$.
Since integrating out at tree level is the same as solving the classical equations of motion and plugging back into the action, and since we need to specify boundary conditions to solve the equations of motion, the fact that these are not invariant implies that the effective action is not invariant either.\footnote{This is to be contrasted with what happens when we integrate out a full multiplet in a {\em linear} representation of a symmetry: zero is always an invariant boundary condition.}
Beyond tree-level the problem remains; within perturbation theory, higher orders cannot cure this zeroth-order problem.

\section{Concluding remarks and outlook}

Our underlying motivation to explore the questions addressed here is related to a more general and more fundamental question, which we discuss at length in a separate paper: for relativistic theories at zero temperature, under what conditions does having a finite charge density imply that the associated symmetry is spontaneously broken?

\vskip4pt
For free bosons, the answer is provided by the phenomenon of Bose--Einstein condensation: the ground state at finite charge density exhibits SSB. Notably this is not the case for free fermions: the ground state at finite density breaks Lorentz boosts, but does not break translations, rotations, or the particle number $U(1)$ symmetry.

\vskip4pt
Beyond free theories, it becomes harder to answer the question in generality. For fermions, {\em if}  Fermi liquid theory applies, then the situation is qualitatively the same as for free fermions \cite{Landau}. For bosons---in particular for scalars---at tree level one finds that a finite density implies SSB, simply because the charge density operator is bilinear in a complex field $\Phi$, so that for example $J_\mu = i \, \Phi^* \overset{\leftrightarrow}{\partial}_\mu \Phi$, and so it is impossible to have nonzero $J^0$ with vanishing $\Phi$. 

\vskip4pt
However, beyond tree-level, in principle it appears to be possible to have a nonzero {\em expectation value} for $J^0$ while having a vanishing one for $\Phi$:
\be 
\langle \Phi^* \overset{\leftrightarrow}{\partial}_0 \Phi \rangle \neq 0 \ , \qquad \langle \Phi \rangle = 0 \; . \label{no SSB}
\ee
The approach that we have taken
 has SSB  built-in from the start---we are deriving the effective action for the gapless Goldstone, which is a consistent procedure only in the case of SSB. Still, the results of Sec.~\ref{SSB section} and \ref{low densities} provide circumstantial evidence that, at least at one loop, the possibility outlined by eq.~\eqref{no SSB} is not realized. 
Indeed, consider the positive  $m^2$ case (when $m^2 <0$, the symmetry is always spontaneously broken, even at vanishing charge density). On the one hand, Sec.~\ref{SSB section} shows that our SSB assumption is consistent only for large enough chemical potentials, $\mu^2 \ge m^2_{\rm pole}$. On the other hand, Sec.~\ref{low densities} shows that for $\mu$ approaching that lower bound from above, the charge density goes to zero. This strongly suggests that the charge density remains zero when $\mu$ is below that bound, that is, when we have no SSB.

\vskip4pt
We tackle the question head-on, without assuming SSB, in forthcoming work.

\paragraph{Acknowledgements}
We are grateful to Paolo Creminelli, Gabriel Cuomo, Luca Delacr\'etaz, Walter Goldberger and Riccardo Rattazzi for useful discussions. AN would also like to thank Paolo Creminelli for earlier collaboration on related topics (unpublished work).
Our work is partially supported by the US DOE (award number DE-SC011941) and by the Simons Foundation (award number 658906).


\appendix

\section{Dimensional regularization integrals}\label{DR integrals}

Here we describe the evaluation of the integral~\eqref{main integral} in dimensional regularization. The starting point is the integral
\be\label{appendix integral}
\Delta \Gamma^{\rm 1 \, loop}_{\rm full}[\phi]  =  \frac{i}{2} \int \dd^4 x \int \frac{\dd^4 p }{(2\pi)^4} \log{\left(p^2-m^2_{\rm eff} - 4\di_\mu \phi \di_\nu \phi \, \frac{p^\mu p^\nu}{p^2} \right)} \; ,
\ee
for generic (but constant) $\di_\mu \phi$ and $m^2_{\rm eff}$. The $\di \phi \di \phi$ structure inside the log prevents us from using spherical coordinates, so instead we will expand in powers of $\di_\mu \phi$, compute the individual integrals, and then try to resum the series. Notice that, as long as $m^2_{\rm eff} \neq 0$, the integral \eqref{appendix integral} is very convergent in the IR, and so the IR-ambiguous  $p^\mu p^\nu/p^2$ structure is harmless.  

\vskip4pt
From expanding the log in powers of $\partial\phi$, we have
\be
\log\left(p^2-m^2_{\rm eff}  - \di_\mu \phi \di_\nu \phi\frac{p^\mu p^\nu}{p^2} \right)  =   \log{\big(p^2-m^2_{\rm eff}\big) - \sum_{n=1}^\infty \frac {4^n} n \di_{\mu_1} \phi \di_{\nu_1}\phi \cdots \di_{\mu_n} \phi \di_{\nu_n} \phi   \frac{ p^{\mu_1} p^{\nu_1} \cdots p^{\mu_n}p^{\nu_n}  }{p^{2n} (p^2- m^2_{\rm eff})^n}} \; . \label{expand}
\ee
Upon integrating in $p$, one can replace the product of momenta $p^\mu$ in the numerator with a totally symmetrized combination of $n$ $\eta^{\mu\nu}$ tensors. The correct normalization factor in $d$ dimensions is \cite{Collins:1984xc}
\be
\frac{p^{\mu_1} p^{\nu_1} \cdots p^{\mu_n}p^{\nu_n}  }{p^{2n} } \longrightarrow  \frac{\Gamma(d/2) \Gamma(n+1/2)}{\Gamma(1/2) \Gamma(n+d/2)} \, \eta^{(\mu_1 \nu_1} \dots \eta^{\mu_n \nu_n)}\,.
\ee
Thus, the right-hand side of~\eqref{expand} can be replaced by
\be
\log\big(p^2-m^2_{\rm eff}\big)  - \frac{\Gamma(d/2)}{\Gamma(1/2) }\sum_{n=1}^\infty \frac 1 n (4X)^n  \,  \frac{\Gamma(n+1/2)}{\Gamma(n+d/2)} \frac{1 }{(p^2- m^2_{\rm eff})^n} \,.
\ee
Or, upon Wick rotating, we can write
\be
\log{\big(p_E^2+m^2_{\rm eff}\big)} - \frac{\Gamma(d/2)}{\Gamma(1/2) }\sum_{n=1}^\infty \frac{(-4X)^n}{n}  \,  \frac{\Gamma(n+1/2)}{\Gamma(n+d/2)} \frac{1 }{(p_E^2+ m^2_{\rm eff})^n} \; . \label{euclidean}
\ee
In order to evaluate the integral of this expression,
we need the standard dim-reg integrals 
\be
\label{eq:int1}
I_n \equiv \mu_{\rm MS}^{4-d} \int \frac{\dd^d p_E }{(2\pi)^d}\frac{1 }{(p_E^2+ m^2_{\rm eff})^n} = \frac{(m^2_{\rm eff})^{2-n}}{16\pi^{2}}\left(\frac{m^2_{\rm eff}}{4\pi \mu_{\rm MS}^2 }\right)^{d/2-2} \frac{\Gamma(n-d/2)}{\Gamma(n)}\,, 
\ee
for positive integer $n$, as well as the integral
\be
\label{eq:int2}
I_{\rm log} \equiv \mu_{\rm MS}^{4-d} \int \frac{\dd^d p_E }{(2\pi)^d} \log \big(p_E^2+m^2_{\rm eff}\big) = -\frac{\di}{\di n}I_n \bigg|_{n \to 0} = -  \frac{m^4_{\rm eff}}{16\pi^{2}}\left(\frac{m^2_{\rm eff}}{4\pi \mu_{\rm MS}^2 }\right)^{d/2-2} \Gamma(-d/2) \; ,
\ee
where $\mu_{\rm MS}$ denotes the minimal subtraction renormalization scale.
Of all these, only $I_{\rm log}$, $I_1$, and $I_2$ diverge for $d \to 4 $. 

\vskip4pt
Then, to compute~\eqref{appendix integral} we expand out the log using~\eqref{expand}, perform the integrals at each order using~\eqref{eq:int1} and~\eqref{eq:int2} and then resum the series to obtain
\be
\begin{aligned}
\Delta \Gamma^{\rm 1 \, loop}_{\rm full}[\phi]   &= \int \dd^4 x \bigg( -\frac{1}{32\pi^2 (d-4)}  \big( m_{\rm eff}^4 + 2 m_{\rm eff}^2 X + 2 X^2 \big)  \\
&\hspace{.4cm} + \frac{1}{384 \pi^2 } \bigg[  9 m_{\rm eff}^4 +18 m_{\rm eff}^2 X +10 X^2 - 6(m_{\rm eff}^4 + 2 m_{\rm eff}^2 X +2 X^2) \log(m_{\rm eff}^2/\bar \mu^2)     \\
&\hspace{2.2cm} - 10 \frac{X^3}{m_{\rm eff}^2}\, {}_3F_2\left[\begin{array}{c}
1,~~1,~~\sfrac{7}{2}\\
4,~~5
\end{array}\left\lvert -\frac{4X}{m_{\rm eff}^2}\right.\right] \bigg]\bigg)
\; ,
\end{aligned}
\ee
where as usual $\bar\mu^2 = 4\pi \e^{-\gamma_E} \mu_{\rm MS}^2$, where $\gamma_E$ is the Euler--Mascheroni constant and $_3 F_2$ is a generalized hypergeometric function. This is precisely the expression~\eqref{one loop} used in the main text.


\clearpage
\phantomsection
\addcontentsline{toc}{section}{References}
\bibliographystyle{utphys}
\providecommand{\href}[2]{#2}\begingroup\raggedright\endgroup

\end{document}